\documentclass[aps,prd,groupedaddress,graphicx,nofootinbib]{revtex4}
\usepackage{amsmath,amssymb,graphics,graphicx,color,epsf}
\usepackage{subfigure}
\usepackage[scanall]{psfrag}  
\usepackage{tikz}

\begin{document}

\title{Type I Hilltop Inflation and the Refined Swampland Criteria}

\author{Chia-Min Lin}
\affiliation{Fundamental Education Center, National Chin-Yi University of Technology, Taichung 41170, Taiwan}


\date{Draft \today}

\begin{abstract}
In this paper, I show that type I hilltop inflation under the current observational constaints could have $M_P^2 \frac{V^{\prime\prime}}{V} \sim -\mathcal{O}(10^{-2})$ in the parameter space which is in tension with the refined swampland criteria. On the other hand, $M_P^2 \frac{V^{\prime\prime}}{V} \sim -\mathcal{O}(1)$ can be achieved if type I hilltop inflation on the brane is considered.
\end{abstract}
\maketitle
\large
\baselineskip 18pt
\section{Introduction}

There are many inflation models which are compatible with current observational data \cite{Akrami:2018odb}, while they may or may not be compatible with string theory. The landscape of string theory is large, however it is thought to be surrounded by a vast swampland of effective field theories which are not compatible with string theory. This provides a way to investigate whether the idea of swampland conjectures \cite{Ooguri:2006in,Obied:2018sgi,Ooguri:2016pdq,Agrawal:2018own,Ooguri:2018wrx} are in tension with the idea of cosmic inflation, namely which models of inflation are (not) in the swampland. 
If eventually string theory with the idea of swampland conjectures is proven to be correct, inflation models in the swampland are ruled out. 
On the other hand, if a particular inflation model in the swampland is observationally confirmed, it may have negative consequences 
for string theory itself.  It has been proposed that if an effective field theory is embedded consistently in string theory, it satisfies two swampland criteria \cite{Ooguri:2006in,Obied:2018sgi}.
\begin{itemize}
\item \textbf{The distance conjecture:}
\begin{equation}
\frac{\Delta \phi}{M_P} < \mathcal{O}(1),
\label{eq1}
\end{equation}
\item \textbf{The de Sitter conjecture:}
\begin{equation}
M_P \frac{|V^\prime|}{V}>c \sim \mathcal{O}(1),
\label{eq2}
\end{equation}
\end{itemize}
The distance conjecture states that scalar field excursion in reduced Planck units in field space are bounded from above \cite{Ooguri:2016pdq} and the de Sitter conjecture states that the slope of the scalar field potential satisfies a lower bound whenever $V>0$ \cite{Agrawal:2018own}. More importantly, both conjectures state that the bounds are of $\mathcal{O}(1)$. The cosmological consequences of these swampland criteria is investigated in \cite{Dvali:2018fqu, Achucarro:2018vey, Garg:2018reu, Lehners:2018vgi, Kehagias:2018uem, Dias:2018ngv, Colgain:2018wgk, Matsui:2018bsy, Ben-Dayan:2018mhe, Damian:2018tlf, Kinney:2018nny, Cicoli:2018kdo, Akrami:2018ylq, Marsh:2018kub, Brahma:2018hrd, Das:2018hqy, Wang:2018duq, Han:2018yrk, Visinelli:2018utg, Brandenberger:2018xnf, Brandenberger:2018wbg, Heisenberg:2018rdu, Gu:2018akj, Heisenberg:2018yae, Brandenberger:2018fdd, Ashoorioon:2018sqb, Odintsov:2018zai, Motaharfar:2018zyb, Kawasaki:2018daf, Lin:2018kjm, Dimopoulos:2018upl, Das:2018rpg, Banks:2018ypk, Andriot:2018wzk}.

Some modifications have been suggested in \cite{Dvali:2018fqu, Garg:2018reu, Andriot:2018wzk, Murayama:2018lie} which have different implications to particle phyisics and cosmology models. Recently the refined version of the de Sitter conjecture is proposed in \cite{Ooguri:2018wrx}. The refinement is essentially the same as the proposal of \cite{Garg:2018reu}. For a single field, it can be stated as the following:
\begin{itemize}
\item \textbf{Refined de Sitter conjecture:}
\begin{equation}
M_P \frac{|V^\prime|}{V}>c \sim \mathcal{O}(1)\mbox{ or } M_P^2 \frac{V^{\prime\prime}}{V} < -c^\prime \sim -\mathcal{O}(1).
\end{equation}
\end{itemize}
The refined version is weaker and allows a scalar field with a potentail maxima, namely a hilltop to exist. Therefore it not only evades the problems of the local maximum in the Higgs and the QCD axion potential which are contradictory with Eq.~(\ref{eq2}) \cite{Denef:2018etk, Choi:2018rze, Murayama:2018lie, Hamaguchi:2018vtv} but also hilltop inflation \cite{Boubekeur:2005zm, Kohri:2007gq}. Some cosmological consequences of the refined swampland conjecture have been discussed in \cite{Wang:2018kly, Fukuda:2018haz, Garg:2018zdg, Park:2018fuj}.

Slow-roll inflation is usually characterized by the slow-roll parameters:
\begin{eqnarray}
\epsilon &\equiv& \frac{M_P^2}{2} \left( \frac{V^\prime}{V} \right)^2 > \frac{1}{2}c^2,   \\
\eta &\equiv& M_P^2 \frac{V^{\prime\prime}}{V} < -c^\prime.
\end{eqnarray}
Hilltop inflation is a small field inflation model, therefore the distance conjecture is automatically satisfied. Interestingly, the alternative condition for the de Sitter conjecture even prefers a convex potential which is the character of a hilltop inflation. However, if $c \sim c^\prime \sim \mathcal{O}(1)$, (conventional) slow-roll inflation does not happen. I will show that if we can tune the parameter to $c^\prime \sim \mathcal{O}(10^{-2})$, type I hilltop inflation would be a viable model.

Another way to satisfy the refined swampland criteria is to consider inflation on the brane \cite{Brahma:2018hrd, Lin:2018kjm}, because the Hubble parameter is enhanced during inflation due to the modification of the Friedmann equation. Since the Hubble parameter effectively provides an "friction term" for the equation of motion of the inflaton field, slow-rolling is enhanced. This mechanism can also be applied to hilltop inflation. This case will be considered in Section \ref{brane}.

\section{Type I Hilltop Inflation}
The inflaton field $\phi$ of a hilltop quartic model has a potential of the form
\begin{equation}
V=V_0-\lambda \phi^4.
\end{equation}
When compared with the lastest Planck satellite result \cite{Akrami:2018odb}, this model can fit the data very well. However, since the slow roll parameter $\eta$ of the model vanishes at the top of the hill when $\phi=0$, this model is incompatible with the refined swampland criteria if $V_0$ is nonzero.
In this paper, I consider the hilltop inflation with the potential
\begin{equation}
V(\phi)=V_0-\frac{1}{2}m^2 \phi^2 - \lambda \frac{\phi^p}{M_P^{p-4}} \equiv V_0 \left(1+\frac{1}{2}\eta_0 \frac{\phi^2}{M_P^2}\right) - \lambda \frac{\phi^p}{M_P^{p-4}} ,
\end{equation}
where
\begin{equation}
\eta_0 \equiv -\frac{m^2 M_P^2}{V_0}.
\end{equation}
This model was originally considered as the first type of hilltop inflation in \cite{Kohri:2007gq} where three types of hilltop inflation models are proposed and analyzed by the author and collaborators. Therefore in this paper, I call it a type I hilltop inflation.

In order to achieve slow-roll inflation, we need $V \simeq V_0$ giving
\begin{eqnarray}
\frac{V^\prime}{V}&=& \eta_0 \frac{\phi}{M_P^2}-p \lambda \frac{\phi^{p-1}}{V_0 M^{p-4}_P}  \\
\frac{V^{\prime \prime}}{V} &=& \frac{\eta_0}{M_P^2}-p(p-1)\lambda \frac{\phi^{p-2}}{V_0 M^{p-4}_P} \label{eta}.
\end{eqnarray}
From Eq.~(\ref{eta}), we can see that $\eta_0$ corresponds to the value of the slow-roll parameter $\eta$ at the top of the potential hill, namely at $\phi=0$. The field value increases during inflation and $\eta$ becomes more negative than $\eta_0$. Therefore when we compare $M_P^2V^{\prime\prime}/V$ with $c^\prime$ for the refined swampland criteria, we should focus on $\eta_0$ instead of $\eta$. The hilltop quartic model corresponds to $\eta_0=0$.
The number of e-folds is given by
\begin{equation}
N=\frac{1}{M_P^2}\int^{\phi(N)}_{\phi_e}\frac{V}{V^\prime}d\phi,
\label{nnn}
\end{equation}
where $\phi_e$ is the field value at the end of inflation and $\phi(N)$ is the field value when there are $N$ e-folds to the end of inflation.    
Eq.~(\ref{nnn}) can be integrated analytically in this model to obtain
\begin{eqnarray}
\left( \frac{\phi}{M_P} \right)^{p-2} &=& \left( \frac{V_0}{M_P^4} \right) \frac{\eta_0 e^{(p-2)\eta_0 N}}{\eta_0 x+ p \lambda (e^{(p-2)\eta_0 N}-1)}    \\
x &\equiv& \left( \frac{V_0}{M_P^4} \right) \left( \frac{M_P}{\phi_e} \right)^{p-2}, \label{x}
\end{eqnarray}
leading to the predictions for the spectrum of the primodial density perturbation $P_\zeta$, the spectral index $n_s$, and the running spectral index $\alpha$ as 
\begin{eqnarray}
P_\zeta&=&\frac{1}{12\pi^2}\left( \frac{V_0}{M_P^4} \right)^{\frac{p-4}{p-2}}e^{-2\eta_0 N}\frac{[p\lambda (e^{(p-2)\eta_0 N}-1)+\eta_0 x]^{\frac{2p-2}{p-2}}}{\eta_0^{\frac{2p-2}{p-2}}(\eta_0 x-p \lambda)^2}   \\ \label{spectrum}
n_s-1 &=&2 \eta= 2\eta_0 \left[1-\frac{\lambda p (p-1) e^{(p-2)\eta_0 N}}{\eta_0 x + p \lambda (e^{(p-2)\eta_0 N}-1)} \right] \label{index}  \\
\alpha &=& 2 \eta_0^2 \lambda p(p-1)(p-2) \frac{e^{(p-2)\eta_0 N}(\eta_0 x -p\lambda)}{[\eta_0 x + p \lambda(e^{(p-2)\eta_0 N}-1)]^2}. \label{running}
\end{eqnarray}

From Eq.~(\ref{eta}), we can see that $\eta$ becomes more and more negtive during inflation when the field value of $\phi$ increases and inflation ends when $\eta=-1$ at $N=0$. From Eq.~(\ref{index}), this implies 
\begin{equation}
x=\frac{\lambda p (p-1)}{\eta_0 + 1}. \label{xlam}
\end{equation} 
By using the definition of $x$ from Eq.~(\ref{x}), we can obtain
\begin{equation}
\frac{V_0}{M_P^4}=\frac{\lambda p (p-1)}{\eta_0+1}\left(\frac{\phi_e}{M_P} \right)^{p-2}. \label{scale}
\end{equation}
This is in accordance with the well known fact that a low scale slow-roll inflation is a small field inflation. In the following, I consider two cases where $p=4$ and $p=6$ as examples.

\section{Case $p=4$}
\begin{figure}[t]
\centering
\includegraphics[width=0.7\columnwidth]{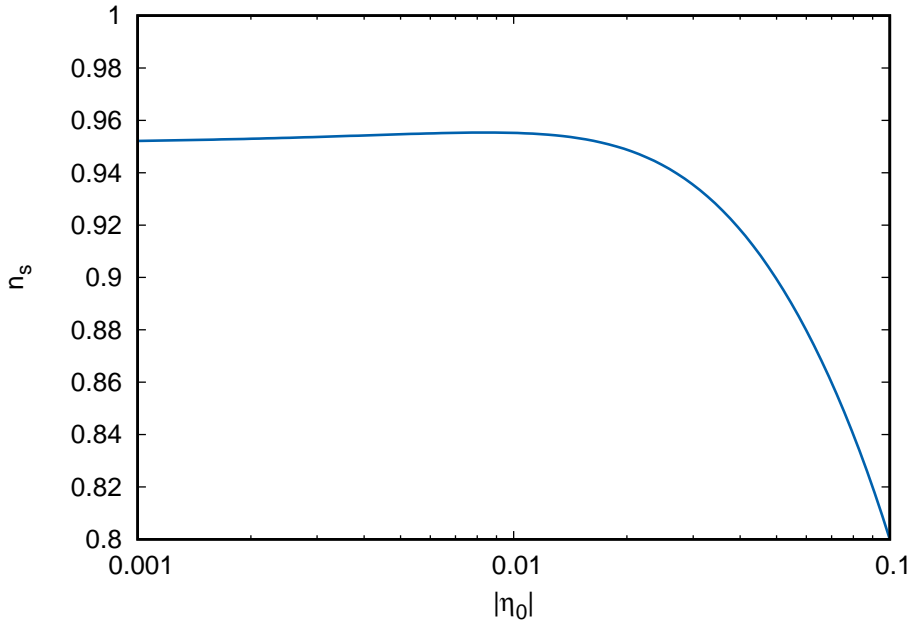}
 \caption{The spectral index $n_s$ as a function of $|\eta_0|$ for $p=4$.}
\label{fig1}
\end{figure}

For the case $p=4$, from Eq.~(\ref{index}), by fixing $N=60$ for the horizon exit of the CMB scale I obtain
\begin{equation}
n_s=1+2\eta_0 \left[1-\frac{3 e^{120 \eta_0}(\eta_0+1)}{3 \eta_0+(e^{120\eta_0}-1)(\eta_0+1)} \right].
\end{equation}
It is interesting to note that this prediction only depends on $\eta_0$. The spectral index $n_s$ as a function of $|\eta_0|$ is given in Fig.~\ref{fig1}. This can be compared with the Planck result $n_s \sim 0.96$ \cite{Akrami:2018odb}. We can see from the figure that $|\eta_0|$ can be as large as $|\eta_0| \sim 0.01$ and the spectral index is still within the experimental bounds.

\begin{figure}[t]
\centering
\includegraphics[width=0.7\columnwidth]{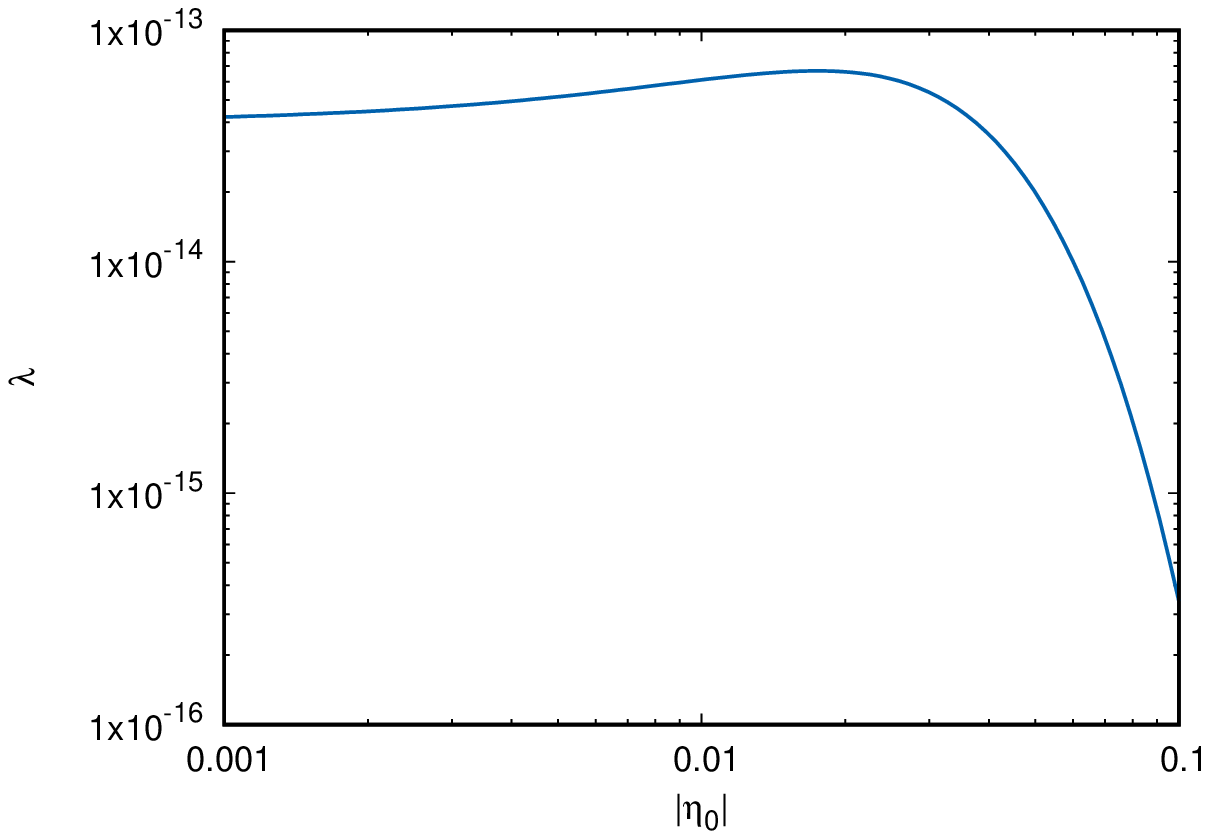}
 \caption{$\lambda$ as a function of $|\eta_0|$ for $p=4$.}
\label{fig2}
\end{figure}

From Eqs.~(\ref{spectrum}) and (\ref{xlam}), by imposing CMB normalization $P_\zeta \sim (5 \times 10^{-5})^2$, we obtain
\begin{equation}
\lambda = (2.96 \times 10^{-7})\times \frac{e^{120 \eta_0}\times \eta_0^3\left(\frac{12\eta_0}{\eta_0+1}-4\right)^2}{\left[ 4(e^{120\eta_0}-1)+\frac{12 \eta_0}{\eta_0+1} \right]^3}.
\end{equation}
This result also only depends on $\eta_0$. I plot $\lambda$ as a function of $|\eta_0|$ in Fig.~\ref{fig2}.

\begin{figure}[t]
\centering
\includegraphics[width=0.7\columnwidth]{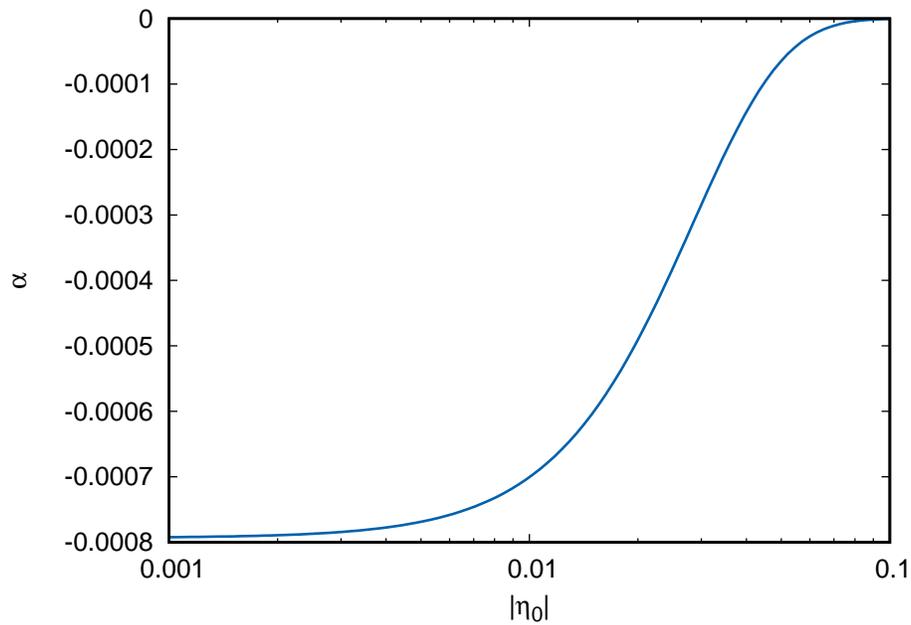}
 \caption{$\alpha$ as a function of $|\eta_0|$ for $p=4$.}
\label{fig3}
\end{figure}

The running spectral index $\alpha$ can be obtained from Eqs.~(\ref{running}) and (\ref{xlam}) as
\begin{equation}
\alpha = 48 \eta_0^2 \frac{e^{120\eta_0}\left( \frac{12\eta_0}{\eta_0+1}-4\right)}{\left[\frac{12\eta_0}{\eta_0+1}+4(e^{120\eta_0}-1)\right]^2}.
\end{equation}
The running specrtral index for $p=4$ as a function of $\eta_0$ is plotted in Fig.~\ref{fig3}.
This is safely within the experimental bound $|\alpha|<0.01$ \cite{Akrami:2018odb}. 

\section{Case $p=6$}
\begin{figure}[t]
\centering
\includegraphics[width=0.7\columnwidth]{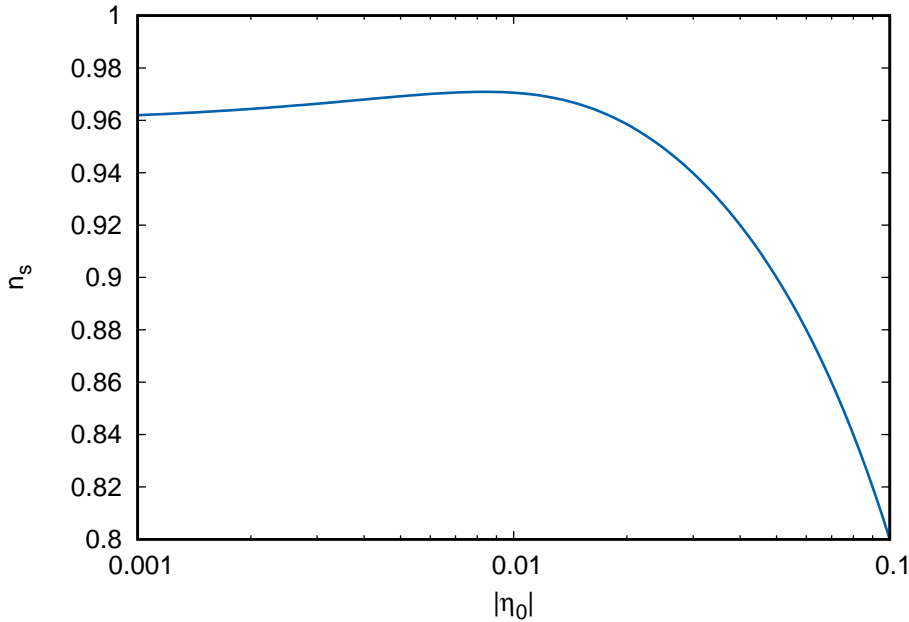}
 \caption{The spectral index $n_s$ as a function of $|\eta_0|$ for $p=6$.}
\label{fig4}
\end{figure}

For the case $p=6$, from Eq.~(\ref{index}), by fixing $N=60$ for the horizon exit we obtain
\begin{equation}
n_s=1+2\eta_0 \left[1-\frac{5 e^{240 \eta_0}(\eta_0+1)}{5 \eta_0+(e^{240\eta_0}-1)(\eta_0+1)} \right].
\end{equation}
The spectral index $n_s$ as a function of $|\eta_0|$ is given in Fig.~\ref{fig4}. This fits the experimental result $n_s \sim 0.96$ perfectly well for $|\eta_0| \lesssim 0.01$.

\begin{figure}[t]
\centering
\includegraphics[width=0.7\columnwidth]{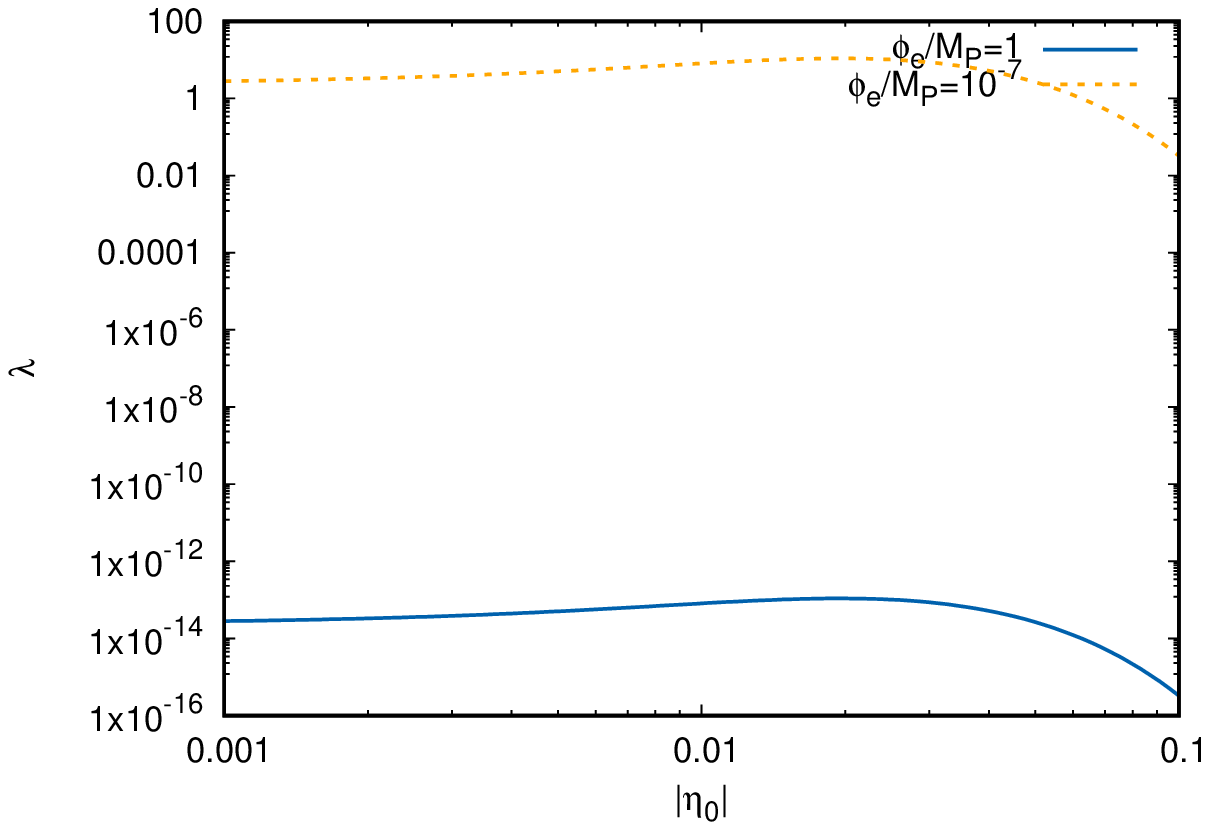}
 \caption{$\lambda$ as a function of $|\eta_0|$ for $p=6$.}
\label{fig5}
\end{figure}

\begin{figure}[t]
\centering
\includegraphics[width=0.7\columnwidth]{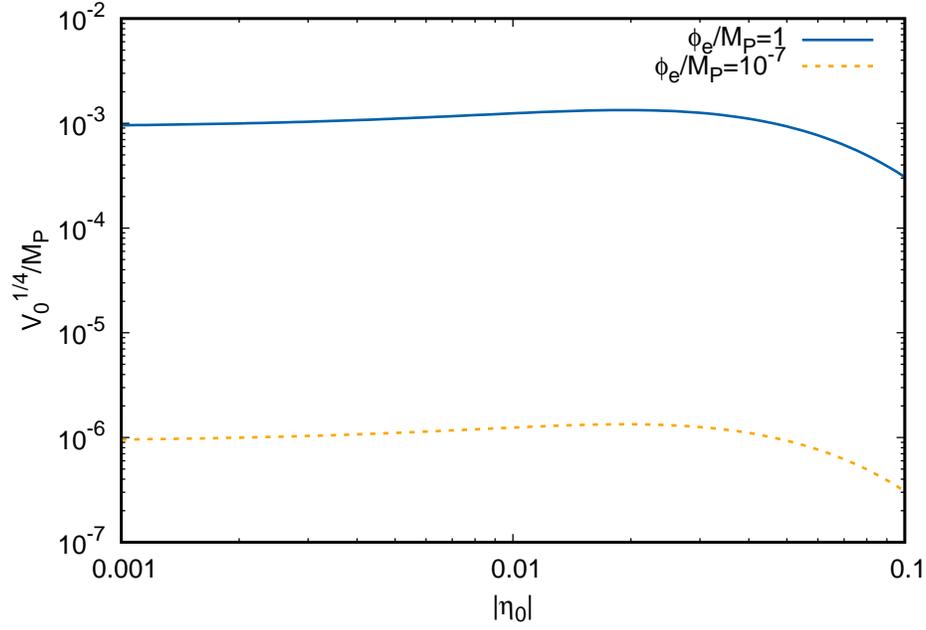}
 \caption{$V_0^{1/4}$ as a function of $|\eta_0|$ for $p=6$.}
\label{fig6}
\end{figure}

From Eqs.~(\ref{spectrum}), (\ref{xlam}) and (\ref{scale}), by imposing CMB normalization $P_\zeta \sim (5 \times 10^{-5})^2$, we can obtain
\begin{equation}
\lambda = (2.96 \times 10^{-7})\times \frac{e^{120 \eta_0}\times \eta_0^{\frac{5}{2}}\left(\frac{30\eta_0}{\eta_0+1}-6\right)^2}{\left[ 6(e^{240\eta_0}-1)+\frac{30 \eta_0}{\eta_0+1} \right]^{\frac{5}{2}}}\frac{M_P^2}{\phi_e^2}\left( \frac{\eta_0+1}{30} \right)^{\frac{1}{2}}.
\end{equation}
This result depends not only on $\eta_0$ but also on the inflaton field value at the end of inflation $\phi_e$. I plot $\lambda$ as a function of $|\eta_0|$ in Fig.~\ref{fig5} for $\phi_e=M_P$ and $\phi_e=10^{-7}M_P$. Here $\phi_e=M_P$ should be regarded as the upper bound for the field value. As can be seen in the plot, by decreasing $\phi_e$, $\lambda$ increases. It is even possible to obtain $\lambda \sim \mathcal{O}(1)$ if $\phi_e$ is small enough. Because of the relation between $\phi_e$ and $V_0$ from Eq.~(\ref{xlam}), one may wonder whether such a small field value could result in a unacceptable small inflation scale $V_0$. Therefore $V_0^{1/4}$ is plotted in Fig.~\ref{fig6}. As can be seen in the plot, $V_0^{1/4}$ in this range of $\phi_e$ is still much larger than the requirement $V_0^{1/4} \gtrsim T_R \gtrsim \mbox{MeV} \sim 10^{-21} M_P$ from successful Big Bang Nucleosynthesis (BBN) \cite{deSalas:2015glj}.

\begin{figure}[t]
\centering
\includegraphics[width=0.7\columnwidth]{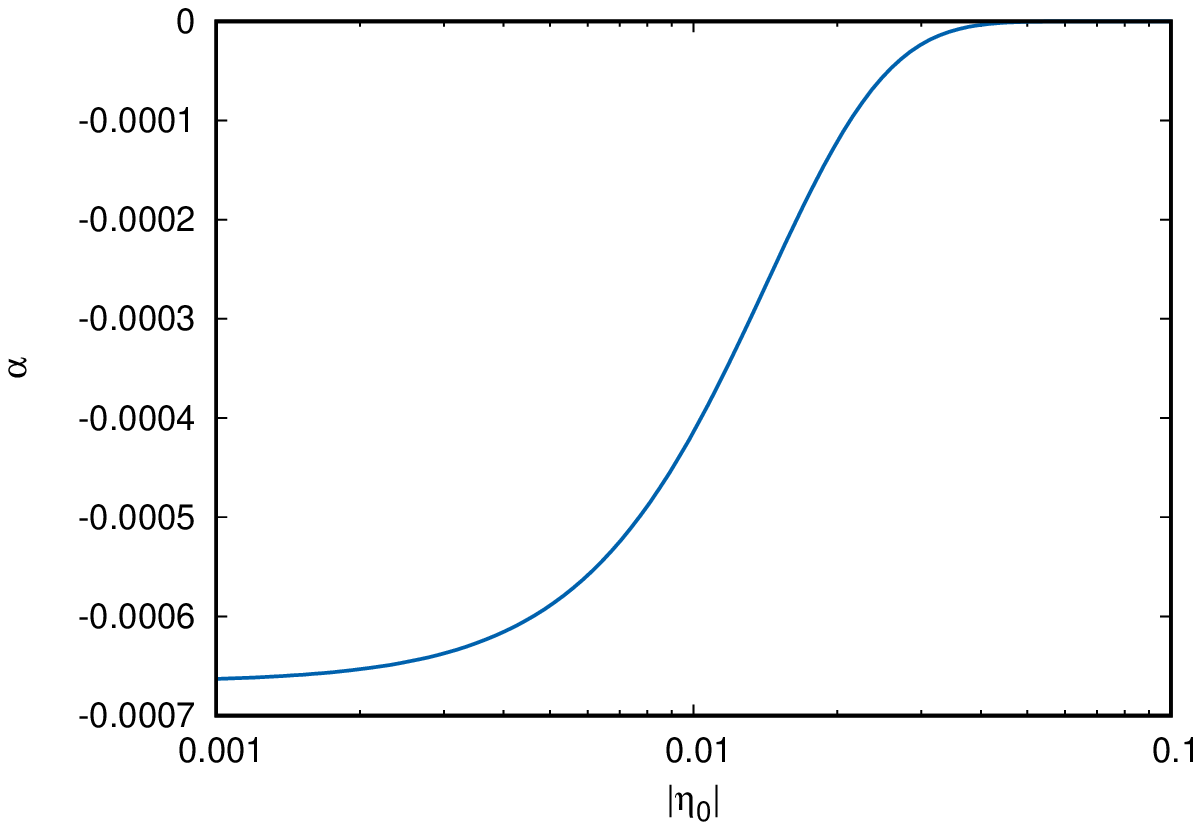}
 \caption{$\alpha$ as a function of $|\eta_0|$ for $p=6$.}
\label{fig7}
\end{figure}

The running spectral index $\alpha$ can be obtained from Eqs.~(\ref{running}) and (\ref{xlam}) as
\begin{equation}
\alpha = 240 \eta_0^2 \frac{e^{240\eta_0}\left( \frac{30\eta_0}{\eta_0+1}-6\right)}{\left[\frac{30\eta_0}{\eta_0+1}+6(e^{240\eta_0}-1)\right]^2}.
\end{equation}
The running specrtral index for $p=6$ as a function of $\eta_0$ is plotted in Fig.~\ref{fig7}.
This is again safely within the experimental bound $|\alpha|<0.01$.

\section{Type I hilltop inflation on the brane}
\label{brane}
For inflation on the brane, the slow-roll parameters are modified into \cite{Maartens:1999hf}
\begin{eqnarray}
\epsilon &\equiv& \frac{M_P^2}{2} \left( \frac{V^\prime}{V} \right)^2 \frac{1}{\left( 1+\frac{V}{2\Lambda} \right)^2}\left( 1+\frac{V}{\Lambda} \right), \label{braneepsilon} \\
\eta &\equiv& M_P^2 \left( \frac{V^{\prime\prime}}{V} \right)\left( \frac{1}{1+\frac{V}{2 \Lambda}} \right) \label{braneeta},
\end{eqnarray}
where $\Lambda$ provides a relation between the four-dimensional Planck scale $M_4$ and five-dimensional Planck scale $M_5$ through
\begin{equation}
M_4=\sqrt{\frac{3}{4\pi}} \left( \frac{M_5^2}{\sqrt{\Lambda}} \right) M_5,
\end{equation}
where $M_P=M_4/\sqrt{8\pi}$.

If $V \gg \Lambda$, the slow-roll parameters are given by
\begin{eqnarray}
\epsilon &=& \frac{M_P^2}{2} \left( \frac{V^\prime}{V} \right)^2 \frac{1}{\left( \frac{V}{4\Lambda} \right)},  \\
\eta &=& M_P^2 \left( \frac{V^{\prime\prime}}{V} \right)\frac{1}{\left(\frac{V}{2 \Lambda}\right)} ,
\end{eqnarray}
The number of e-folds is 
\begin{equation}
N=\int^{\phi_i}_{\phi_e} \left( \frac{V}{V^\prime} \right) \left( \frac{V}{2 \Lambda} \right) d\phi.
\end{equation}
The spectrum is 
\begin{equation}
P_R=\frac{1}{12\pi^2}\frac{V^3}{V^{\prime 2}} \left(\frac{V}{2 \Lambda} \right)^3. 
\end{equation} 
For hilltop inflation, we have $V \simeq V_0$. Therefore from the above equations, we can see that the result of the previous sections can apply simply by replacing $V_0$ with $V_0^2/\Lambda$. The only exception is the expression for $\epsilon$, but it does not matter because $\epsilon$ is very small and negligible in hilltop inflation due to the fact that it is a small field inflation model. This also implies the predicted tensor-to-scalar ratio in general is very small. We can see from Eq.~(\ref{braneeta}) that we can achive $\eta_0=-0.01$ by having $M^2_P V^{\prime\prime}/V=-1$ and $V/\Lambda=200$. Therefore type I hilltop inflation on the brane could evade the refined swampland criteria completely.

\section{Conclusion and Discussion}
\label{con}
The refined swampland criteria have rescued hilltop inflation on the brane with its topological eternal inflation \cite{Vilenkin:1994pv, Linde:1994hy, Linde:1994wt} from the swampland. In this paper, I have shown that type I hilltop inflation is compatible with the Planck data even when $|\eta_0|$ as large as $|\eta_0|=0.01$. This means we have $M_P^2 V^{\prime\prime}/V < -0.01$ when the field value is non-zero during inflation. However, this is in strong tension with the proposal $c^\prime \sim 1$, although the precise value of $c^\prime$ may depend on the detailed string construction and it may be slightly deviate from unity along the line of argument given in \cite{Dias:2018ngv}. Thus I conclude that hilltop quartic inflation model is incompatible with the refined swampland criteria and type I hilltop inflation model is in tension with it. On the other hand, type I hilltop inflation on the brane can completely satisfy the refined swampland criteria.  
\section*{Acknowledgement}
This work is supported by the Ministry of Science and Technology (MOST) of Taiwan under grant number MOST 106-2112-M-167-001. CML thanks Kin-Wang Ng for useful discussions.


\begin{thebibliography}{99}

\bibitem{Akrami:2018odb} 
  Y.~Akrami {\it et al.} [Planck Collaboration],
  arXiv:1807.06211 [astro-ph.CO].


\bibitem{Ooguri:2006in} 
  H.~Ooguri and C.~Vafa,
  Nucl.\ Phys.\ B {\bf 766}, 21 (2007)
  doi:10.1016/j.nuclphysb.2006.10.033
  [hep-th/0605264].




\bibitem{Obied:2018sgi} 
  G.~Obied, H.~Ooguri, L.~Spodyneiko and C.~Vafa,
  arXiv:1806.08362 [hep-th].


\bibitem{Ooguri:2016pdq} 
  H.~Ooguri and C.~Vafa,
  Adv.\ Theor.\ Math.\ Phys.\  {\bf 21}, 1787 (2017)
  doi:10.4310/ATMP.2017.v21.n7.a8
  [arXiv:1610.01533 [hep-th]].




\bibitem{Agrawal:2018own} 
  P.~Agrawal, G.~Obied, P.~J.~Steinhardt and C.~Vafa,
  Phys.\ Lett.\ B {\bf 784}, 271 (2018)
  doi:10.1016/j.physletb.2018.07.040
  [arXiv:1806.09718 [hep-th]].


\bibitem{Ooguri:2018wrx} 
  H.~Ooguri, E.~Palti, G.~Shiu and C.~Vafa,
  arXiv:1810.05506 [hep-th].

\bibitem{Dvali:2018fqu} 
  G.~Dvali and C.~Gomez,
  arXiv:1806.10877 [hep-th].

\bibitem{Achucarro:2018vey} 
  A.~Achúcarro and G.~A.~Palma,
  arXiv:1807.04390 [hep-th].

\bibitem{Garg:2018reu} 
  S.~K.~Garg and C.~Krishnan,
  arXiv:1807.05193 [hep-th].

\bibitem{Lehners:2018vgi} 
  J.~L.~Lehners,
  arXiv:1807.05240 [hep-th].

\bibitem{Kehagias:2018uem} 
  A.~Kehagias and A.~Riotto,
  arXiv:1807.05445 [hep-th].

\bibitem{Dias:2018ngv} 
  M.~Dias, J.~Frazer, A.~Retolaza and A.~Westphal,
  arXiv:1807.06579 [hep-th].

\bibitem{Colgain:2018wgk} 
  E.~Ó.~Colgáin, M.~H.~P.~M.~Van Putten and H.~Yavartanoo,
  arXiv:1807.07451 [hep-th].

\bibitem{Matsui:2018bsy} 
  H.~Matsui and F.~Takahashi,
  arXiv:1807.11938 [hep-th].

\bibitem{Ben-Dayan:2018mhe} 
  I.~Ben-Dayan,
  arXiv:1808.01615 [hep-th].

\bibitem{Damian:2018tlf} 
  C.~Damian and O.~Loaiza-Brito,
  arXiv:1808.03397 [hep-th].

\bibitem{Kinney:2018nny} 
  W.~H.~Kinney, S.~Vagnozzi and L.~Visinelli,
  arXiv:1808.06424 [astro-ph.CO].

\bibitem{Cicoli:2018kdo} 
  M.~Cicoli, S.~De Alwis, A.~Maharana, F.~Muia and F.~Quevedo,
  arXiv:1808.08967 [hep-th].

\bibitem{Akrami:2018ylq} 
  Y.~Akrami, R.~Kallosh, A.~Linde and V.~Vardanyan,
  arXiv:1808.09440 [hep-th].

\bibitem{Marsh:2018kub} 
  M.~C.~D.~Marsh,
  arXiv:1809.00726 [hep-th].

\bibitem{Brahma:2018hrd} 
  S.~Brahma and M.~Wali Hossain,
  arXiv:1809.01277 [hep-th].

\bibitem{Das:2018hqy} 
  S.~Das,
  arXiv:1809.03962 [hep-th].

\bibitem{Wang:2018duq} 
  D.~Wang,
  arXiv:1809.04854 [astro-ph.CO].

\bibitem{Han:2018yrk} 
  C.~Han, S.~Pi and M.~Sasaki,
  arXiv:1809.05507 [hep-ph].

\bibitem{Visinelli:2018utg} 
  L.~Visinelli and S.~Vagnozzi,
  arXiv:1809.06382 [hep-ph].

\bibitem{Brandenberger:2018xnf} 
  R.~Brandenberger, R.~R.~Cuzinatto, J.~Fröhlich and R.~Namba,
  arXiv:1809.07409 [gr-qc].

\bibitem{Brandenberger:2018wbg} 
  R.~H.~Brandenberger,
  arXiv:1809.04926 [hep-th].

\bibitem{Heisenberg:2018rdu} 
  L.~Heisenberg, M.~Bartelmann, R.~Brandenberger and A.~Refregier,
  arXiv:1809.00154 [astro-ph.CO].

\bibitem{Gu:2018akj} 
  B.~M.~Gu and R.~Brandenberger,
  arXiv:1808.03393 [hep-th].

\bibitem{Heisenberg:2018yae} 
  L.~Heisenberg, M.~Bartelmann, R.~Brandenberger and A.~Refregier,
  arXiv:1808.02877 [astro-ph.CO].

\bibitem{Brandenberger:2018fdd} 
  R.~Brandenberger, L.~L.~Graef, G.~Marozzi and G.~P.~Vacca,
  arXiv:1807.07494 [hep-th].

\bibitem{Ashoorioon:2018sqb} 
  A.~Ashoorioon,
  arXiv:1810.04001 [hep-th].

\bibitem{Odintsov:2018zai} 
  S.~D.~Odintsov and V.~K.~Oikonomou,
  arXiv:1810.03575 [gr-qc].

\bibitem{Motaharfar:2018zyb} 
  M.~Motaharfar, V.~Kamali and R.~O.~Ramos,
  arXiv:1810.02816 [astro-ph.CO].

\bibitem{Kawasaki:2018daf} 
  M.~Kawasaki and V.~Takhistov,
  arXiv:1810.02547 [hep-th].

\bibitem{Lin:2018kjm} 
  C.~M.~Lin, K.~W.~Ng and K.~Cheung,
  arXiv:1810.01644 [hep-ph].

\bibitem{Dimopoulos:2018upl} 
  K.~Dimopoulos,
  arXiv:1810.03438 [gr-qc].

\bibitem{Das:2018rpg} 
  S.~Das,
  arXiv:1810.05038 [hep-th].

\bibitem{Banks:2018ypk} 
  T.~Banks and W.~Fischler,
  arXiv:1806.01749 [hep-th].

\bibitem{Andriot:2018wzk} 
  D.~Andriot,
  Phys.\ Lett.\ B {\bf 785}, 570 (2018)
  doi:10.1016/j.physletb.2018.09.022
  [arXiv:1806.10999 [hep-th]].

\bibitem{Murayama:2018lie} 
  H.~Murayama, M.~Yamazaki and T.~T.~Yanagida,
  arXiv:1809.00478 [hep-th].













\bibitem{Denef:2018etk} 
  F.~Denef, A.~Hebecker and T.~Wrase,
  Phys.\ Rev.\ D {\bf 98}, no. 8, 086004 (2018)
  doi:10.1103/PhysRevD.98.086004
  [arXiv:1807.06581 [hep-th]].

\bibitem{Choi:2018rze} 
  K.~Choi, D.~Chway and C.~S.~Shin,
  arXiv:1809.01475 [hep-th].



\bibitem{Hamaguchi:2018vtv} 
  K.~Hamaguchi, M.~Ibe and T.~Moroi,
  arXiv:1810.02095 [hep-th].

\bibitem{Boubekeur:2005zm} 
  L.~Boubekeur and D.~H.~Lyth,
  JCAP {\bf 0507}, 010 (2005)
  doi:10.1088/1475-7516/2005/07/010
  [hep-ph/0502047].


\bibitem{Kohri:2007gq} 
  K.~Kohri, C.~M.~Lin and D.~H.~Lyth,
  JCAP {\bf 0712}, 004 (2007)
  doi:10.1088/1475-7516/2007/12/004
  [arXiv:0707.3826 [hep-ph]].

\bibitem{Wang:2018kly} 
  S.~J.~Wang,
  arXiv:1810.06445 [hep-th].

\bibitem{Fukuda:2018haz} 
  H.~Fukuda, R.~Saito, S.~Shirai and M.~Yamazaki,
  arXiv:1810.06532 [hep-th].

\bibitem{Garg:2018zdg} 
  S.~K.~Garg, C.~Krishnan and M.~Z.~Zaz,
  arXiv:1810.09406 [hep-th].

\bibitem{Park:2018fuj} 
  S.~C.~Park,
  arXiv:1810.11279 [hep-ph].



\bibitem{deSalas:2015glj} 
  P.~F.~de Salas, M.~Lattanzi, G.~Mangano, G.~Miele, S.~Pastor and O.~Pisanti,
  Phys.\ Rev.\ D {\bf 92}, no. 12, 123534 (2015)
  doi:10.1103/PhysRevD.92.123534
  [arXiv:1511.00672 [astro-ph.CO]].

\bibitem{Maartens:1999hf} 
  R.~Maartens, D.~Wands, B.~A.~Bassett and I.~Heard,
  Phys.\ Rev.\ D {\bf 62}, 041301 (2000)
  doi:10.1103/PhysRevD.62.041301
  [hep-ph/9912464].

\bibitem{Vilenkin:1994pv} 
  A.~Vilenkin,
  Phys.\ Rev.\ Lett.\  {\bf 72}, 3137 (1994)
  doi:10.1103/PhysRevLett.72.3137
  [hep-th/9402085].

\bibitem{Linde:1994hy} 
  A.~D.~Linde,
  Phys.\ Lett.\ B {\bf 327}, 208 (1994)
  doi:10.1016/0370-2693(94)90719-6
  [astro-ph/9402031].

\bibitem{Linde:1994wt} 
  A.~D.~Linde and D.~A.~Linde,
  Phys.\ Rev.\ D {\bf 50}, 2456 (1994)
  doi:10.1103/PhysRevD.50.2456
  [hep-th/9402115].

\end{thebibliography}
\end{document}